\documentclass{ws-mpla}

\usepackage{psfrag}

\begin{document}

\markboth{Mu-Chun Chen}
{Models of Little Higgs and Electroweak Precision Tests}

\catchline{}{}{}{}{}

\title{Models of Little Higgs and Electroweak Precision Tests
}

\author{\footnotesize Mu-Chun Chen
}

\address{Theoretical Physics Department, Fermi National Accelerator Laboratory\\
MS 106, P.O. Box 500,  Batavia, IL 60510, U.S.A.\\
mcchen@fnal.gov}

\maketitle

\pub{Received (Day Month Year)}{Revised (Day Month Year)}

\begin{abstract}
The little Higgs idea is an alternative to supersymmetry as a solution to 
the gauge hierarchy problem. In this note, I review various 
little Higgs models and their phenomenology with emphases on the precision 
electroweak constraints in these models. 
\end{abstract}
\ccode{14.80.Cp,12.15.Lk,12.60.Cn}

\section{Introduction}	

The Standard Model (SM) requires a Higgs boson to explain the 
generation of fermion and gauge boson masses. The precise electroweak 
(EW) measurements at LEP suggest
that the Higgs boson must be relatively light\cite{LEPEWWG}, with $m_{H} <219$ GeV. 
A light Higgs is also required in order to unitarize the longitudinal 
$W_{L}-W_{L}$ scattering amplitude\cite{Lee:1977eg}. 
In addition, the triviality bound indicates that the SM with a light Higgs can 
be a theory valid all the way up to the Planck scale.  This simple assumption 
of the SM with a single Higgs doublet, however, has the theoretical problem 
that the Higgs boson mass is quadratically sensitive 
to any new physics which may arise at high energy scales. To cancel these 
quadratic divergences due to the SM particles, a set of new states, 
which are related to the SM particles by some symmetry, have to appear 
at the TeV scale\footnote{It has been realized recently in twin 
Higgs models that this needs not be the case\cite{Chacko:2005pe}.}. Little 
Higgs (LH) models are a new approach to stabilizing the mass of the Higgs boson. 
These models have an expanded gauge structure at the TeV scale which 
contains the Standard Model $SU(2)\times U(1)$ electroweak gauge groups.  
They are constructed
such that there are multiple global symmetries that prohibit the Higgs boson from 
obtaining a quadratically divergent mass. It is only when {\it all} these global 
symmetries are broken can a quadratic contribution to the scalar 
potential arises, which is at least at the two loop order.

There is generally a tension between the solution to the gauge hierarchy 
problem and the EW precision fit, nevertheless. The deviations from the 
SM predictions due to the presence of new interactions above the SM 
cutoff scale, $\Lambda$, can be parameterized by a set of higher 
dimensional operators which have been classified in 
Ref.~ \refcite{Buchmuller:1985jz}.   Among these operators, 
the most stringent bounds are those on the coefficients of the 
dim-6 operator, $\frac{1}{\Lambda^{2}}(H^{\dagger}D_{\mu}H)^{2}$, 
which breaks the custodial SU(2) symmetry, and  
$\frac{1}{\Lambda^{2}}(D^{2}H^{\dagger}D^{2}H)$, 
and thus contributes to the S-parameter. These bounds indicate 
that the cutoff scale $\Lambda$ has to be above 5 TeV.  Thus 
the form of possible new Physics, including the little Higgs models, 
which have to appear at the TeV scale to solve the gauge hierarchy 
problem, is severely constrained.

This review is organized as follows. In Sec.~\ref{intro}, I introduce 
the basic idea of the little Higgs models, show explicitly how the 
quadratic divergences are cancelled in various sectors of the littlest 
Higgs model, and briefly review other existing little Higgs models. Sec.~\ref{preew} 
is devoted to the precision electroweak constraints in these models, followed by 
Sec.~\ref{other} in which other issues such as implicit fine-tuning and UV completion are 
briefly discussed. Sec.~\ref{concl} concludes this review.

\section{Little Higgs Models}\label{intro}

The idea of Higgs boson being a pseudo-Goldstone boson  arising from 
the breaking of some approximate global symmetry was proposed\cite{Kaplan:1983fs} 
in the early 80's. Because the Higgs boson mass is generated radiatively, 
this therefore provides a natural way to understand why the Higgs boson 
is so light. In its early realizations, the quadratic contributions 
to the Higgs boson mass arise at one-loop, leading to a Higgs mass that 
is still too heavy because it is only suppressed by the one-loop factor.  
The new ingredient in the little Higgs models is the so-called 
collective symmetry breaking. The idea is to choose the gauge and Yukawa 
interactions in such a way that by turning off some part 
of these interactions, the model has enhanced global symmetries that 
forbid a quadratic contribution to the Higgs mass. 
As a result, the quadratic contributions to the Higgs mass can arise only when 
two or more operators are involved, which can occur only at the two loop 
level or beyond. This leads to a Higgs mass $\mu^{2}$ of the order of 
$\mu^{2} \sim \frac{f^{2}}{16\pi^{2}}$. Note that, 
the logarithmic contributions to 
the Higgs mass can still appear at one-loop. These models have the 
following general structure: 
at the electroweak scale, $v \sim \frac{g^{2}f}{4\pi} \sim 200$ GeV, 
there are one or two Higgs doublets and possibly a few additional 
scalar fields; at the scale $g \cdot f \sim 1$ TeV, there exist 
new gauge bosons and fermions;  above the cut-off scale of the 
non-linear sigma model, $\Lambda \sim 4\pi f \sim 10$ 
TeV, the model becomes strongly interacting.
%
Various realizations of the little Higgs idea are 
described below\footnote{For recent reviews on the little Higgs models, see 
Ref.~\refcite{Schmaltz:2005ky}.}.


\subsection{Original Littlest Higgs-like Models}

\subsubsection{Littlest Higgs Model}

The minimal realization of the little Higgs idea is the littlest 
Higgs model\cite{Arkani-Hamed:2002qy}, 
which is a non-linear sigma model based on $SU(5)/SO(5)$. The 
$SU(5)$ global symmetry in the model 
is broken down to $SO(5)$ by the VEV of the sigma field, 
$\left<\Sigma\right>$, which transforms as an adjoint under $SU(5)$, where, 
\begin{equation}
\bigl< \Sigma \bigr> = \Sigma_{0} = \left(
\begin{array}{ccc}
 & & \mbox{I}_{2 \times 2}\\
 & 1 & \\
 \mbox{I}_{2 \times 2} & &
 \end{array}\right) \; .
 \end{equation}
The sigma field can be expanded around the VEV in terms of 
the Goldstone modes, $\Pi\equiv \pi^{a} X^{a}$, as,
\begin{equation}
 \Sigma  = e^{2i\frac{\Pi}{f}}\Sigma_{0} = \Sigma_{0} + 
2i\frac{\Pi}{f} \Sigma_{0} + .....
\end{equation} 
where $X^{a}$ correspond to the broken SU(5) generators. 
The gauge subgroup is chosen to be  
$\left[ SU(2)\times U(1) \right]_{1} \times \left[ SU(2) 
\times U(1) \right]_{2}$. It is broken down to 
its diagonal subgroup, $\left[ SU(2) \times U(1) \right]_{\mbox{\tiny SM}}$, which is 
identified as the SM gauge group. The kinetic term of the sigma field can be written as
\begin{equation}\label{kin}
\mathcal{L}_{\Sigma} = \frac{1}{8} f^{2} \mbox{Tr}\bigl[(D\Sigma^{\dagger})
(D\Sigma)\bigr] \; ,
\end{equation}
where the covariant derivative is, 
\begin{equation}
D_{\mu} \Sigma = \partial_{\mu} \Sigma - i \sum_{j=1,2} 
\biggl[ g_{j} W_{j}^{a} (Q_{j}^{a} \Sigma + \Sigma Q_{j}^{aT}) 
+ g_{j}^{\prime} B_{j} (Y_{j} \Sigma + \Sigma Y_{j}^{T} ) \biggr] \; .
\end{equation}
The generators of the two $SU(2)$ gauge groups, $Q_{1}^{a}$ and 
$Q_{2}^{a}$ for $(a=1,2,3)$, are,
\begin{equation}
Q_{1}^{a}= \left(\begin{array}{c|c}
\frac{1}{2}\sigma^{a} & \hspace{0.4mm} 0_{\     {2\times 3}} \\
\hline
0_{\      3\times 2} &  0_{\      3 \times 3}
\end{array}\right), \quad
Q_{2}^{a}=\left(\begin{array}{c|c}
0_{\      3\times 3} & \hspace{0.4mm}  
0_{\      3\times 2}\\
\hline
0_{\      2\times 3} & -\frac{1}{2}\sigma^{a\ast}
\end{array}\right) \; ,
\end{equation} 
and the two U(1) generators, $Y_{1}$ and $Y_{2}$,  are,
\begin{equation}
Y_{1} = \frac{1}{10}\mbox{diag}\left(-3,-3,2,2,2\right),
\quad
Y_{2} = \frac{1}{10}\mbox{diag}\left(-2,-2,-2,3,3\right) \; .
\end{equation}
When the gauge coupling constants $g_{2}$ and $g_{2}^{\prime}$ are 
turned off, a global symmetry $SU(3)_{1}$, which acts on the last 
three indices, is restored,
\begin{equation}
\left(\begin{array}{cc|l}
\quad & \; & { } \\
\hline 
& &  SU(3)_{1}
\end{array}\right) \; .
\end{equation}
Similarly, when $g_{1}$ and $g_{1}^{\prime}$ are turned off, 
there is an enhanced $SU(3)_{2}$ global symmetry acting on 
the first three indices,
\begin{equation}
\left(\begin{array}{l|cc}
SU(3)_{2} & &\\
\hline
\quad & \; & { } 
\end{array}\right) \; .
\end{equation}
Each of these two $SU(3)$ symmetries individually forbids a quadratic 
contribution to the Higgs potential. Thus a quadratic contribution 
to the Higgs potential can arise only when both global symmetries 
are broken, which is possible only at the two loop level. 

The 14 Goldstone bosons resulting from the breaking of the 
global symmetry can be decomposed in the following way,
\begin{equation}
14 = 4 \oplus 10 = 1_{0} \oplus 3_{0} \oplus 2_{\pm 1/2} \oplus 3_{\pm 1} \; ,
\end{equation}
where the subscripts in the above equation denote the hypercharges. 
The components $1_{0}$ and $3_{0}$ are eaten and become the 
longitudinal degrees of freedom of the heavy gauge bosons, $A_{H}$, 
$Z_{H}$ and $W_{H}$. In the low energy spectrum, there are 
two scalar fields:  the component $2_{\pm 1/2}$ is identified 
as the complex doublet Higgs boson, $h$, of the SM, while $3_{\pm 1}$ 
is an additional complex $SU(2)_{L}$ triplet Higgs, $\Phi$. In terms of these 
low energy degrees of freedom, the $\Pi$ field can be written as,
\begin{equation}
\Pi = \left(
\begin{array}{ccc}
0 & \frac{h^{\dagger}}{\sqrt{2}} & \Phi^{\dagger}\\
\frac{H}{\sqrt{2}} & 0 & \frac{H^{\ast}}{\sqrt{2}}\\
\Phi & \frac{h^{T}}{\sqrt{2}} & 0
\end{array}\right) \; ,
\end{equation} 
where
\begin{equation}
H = \left(\begin{array}{cc}
h^{+} & h^{0}
\end{array}\right) \; , 
\Phi = \left(
\begin{array}{cc}
\phi^{\scriptscriptstyle{++}} & \frac{\phi^{+}}{\sqrt{2}} 
\\
\frac{\phi^{+}}{\sqrt{2}} & \phi^{0}
\end{array}\right) \; .
\end{equation} 

Expanding the kinetic terms of the $\Sigma$ field in Eq.~(\ref{kin}), 
one finds, 
\begin{eqnarray}
\mathcal{L}_{\Sigma} & \rightarrow & \frac{f^{2}}{8} \mbox{Tr} \biggl|
\sum_{j=1,2} \biggl[
g_{j} W_{j} (Q_{j}\Sigma_{0} + \Sigma_{0} Q_{j}^{T}) + 
g_{j}^{\prime} B_{j} (Y_{j} \Sigma_{0} + \Sigma_{0} Y_{j}^{T}) \biggr] \biggr|^{2}
\\
& \rightarrow & 
\frac{f^{2}}{8} \bigg\{ \biggl[
g_{1}^{2} W_{1} W_{1} - 2 g_{1}g_{2} W_{1} W_{2} - 2 g_{1} g_{2} W_{1} W_{2} 
+ g_{2}^{2} W_{2} W_{2} \biggr]
\nonumber\\
&&\quad
+\frac{1}{5} \biggl[
g_{1}^{\prime 2} B_{1} B_{1} - 2 g_{1}^{\prime} g_{2}^{\prime} 
B_{1} B_{2} + g_{2}^{\prime 2} B_{2} B_{2} 
\biggr] \bigg\} \; .
\nonumber
\end{eqnarray}
This gives the following mass matrices for the gauge bosons,
\begin{eqnarray}
M(W) & = &
\left( \begin{array}{cc}
W_{1} & W_{2} \end{array}\right)
\left(  \begin{array}{cc}
g_{1}^{2} & -g_{1}g_{2}\\
-g_{1}g_{2} & g_{2}^{2}\end{array}\right)
\left(\begin{array}{cc}
W_{1} \\ W_{2}
\end{array}\right)
\\
M(B) & = & 
\left( \begin{array}{cc}
B_{1} & B_{2} \end{array}\right)
\left(  \begin{array}{cc}
g_{1}^{\prime 2} & -g_{1}^{\prime} g_{2}^{\prime}\\
-g_{1}^{\prime} g_{2}^{\prime} & g_{2}^{\prime 2}\end{array}\right)
\left(\begin{array}{cc}
B_{1} \\ B_{2}
\end{array}\right) \; ,
\end{eqnarray}
and the mass eigenstates of these mass matrices are
\begin{eqnarray}
W_{L} & = & s W_{1} + c W_{2}, \quad \mbox{with} \quad M_{W_{L}}=0 \; ,
\\
W_{H} & = & -cW_{1} + s W_{2}, \quad \mbox{with} \quad M_{W_{H}} = \frac{1}{2}\sqrt{g_{1}^{2}
+g_{2}^{2}} f \; ,
\\
B_{L} & = & s^{\prime} B_{1} + c^{\prime} B_{2}, \quad \mbox{with} \quad M_{B_{L}}=0
\; , \\
B_{H} & = & -c^{\prime} B_{1} + s^{\prime} B_{2}, \quad \mbox{with} \quad M_{B_{H}}
= \sqrt{\frac{g_{1}^{\prime 2}+g_{2}^{\prime 2}}{20}} f 
\; ,
\end{eqnarray}
where the mixing angle $s$ is $s=\frac{g_{2}}{\sqrt{g_{1}^{2}+g_{2}^{2}}}$ and 
$s^\prime=\frac{g_{2}}{\sqrt{g_{1}^{2}+g_{2}^{2}}}$ while 
$c=\sqrt{1-s^{2}}$ and $c^\prime=\sqrt{1-s^{\prime 2}}$. 
The two massless eigenstates, $W_{L}$ and $B_{L}$ are identified as the 
weak gauge bosons in the SM. The two massive eigenstates, $W_{H}$ and 
$B_{H}$, are the additional gauge bosons having masses of the order of $f$.
The gauge coupling constants of the unbroken subgroup, $SU(2)_{L}$ and $U(1)_{Y}$, 
are given by,
\begin{equation}
g=\frac{g_{1}g_{2}}{\sqrt{g_{1}^{2}+g_{2}^{2}}}, \;
g^{\prime} = \frac{g_{1}^{\prime}g_{2}^{\prime}}
{\sqrt{g_{1}^{\prime 2} + g_{2}^{\prime 2}}} \; .
\end{equation}
  
\begin{figure}[b]
\centerline{\psfig{file=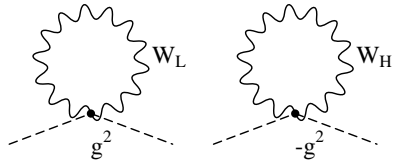,width=2in}}
\vspace*{8pt}
\caption{The cancellation of the quadratic contributions to Higgs mass square at one
loop in the gauge sector.\label{gaugeloop}}
\end{figure}

The quartic couplings of the Higgs boson to the gauge bosons 
arise from the next-to-leading order terms in the expansion in Eq.~(\ref{kin}),
\begin{eqnarray}
\mathcal{L}_{\Sigma} & \rightarrow & \frac{1}{2} \mbox{Tr} \biggl|
\sum_{j=1,2} \biggl[ g_{j} W_{j} (Q_{j}\Pi \Sigma_{0} + \Pi \Sigma_{0} Q_{j}^{T})
+ g_{j}^{\prime} B_{j} (Y_{j}\Pi \Sigma_{0} + \Pi \Sigma_{0} 
Y_{j}^{T})\biggr]\bigg|^{2} 
\\
&\rightarrow&
\frac{1}{4} (g_{1} g_{2} W_{1} W_{2} + 
g_{1}^{\prime} g_{2}^{\prime} B_{1} B_{2} ) H^{\dagger} H + ...
\nonumber\\
& = & \frac{1}{4} \biggl[ g^{2} (W_{L}W_{L}-W_{H} W_{H}) + 
g^{\prime 2} (B_{L}B_{L}-B_{H} B_{H})\biggr] H^{\dagger} H + .....
\nonumber
\end{eqnarray} 
Thus the quartic couplings $H^{\dagger} H W_{L} W_{L}$ and 
$H^{\dagger} H W_{H}W_{H}$ are of equal 
magnitude but opposite signs. The opposite signs 
come about because the $W_{L}$ and $W_{H}$ gauge 
bosons are orthogonal to each other. The cancellation 
of quadratic divergences among diagrams shown in Fig.~\ref{gaugeloop} 
at one loop thus ensues.

\begin{figure}[b]
\centerline{\psfig{file=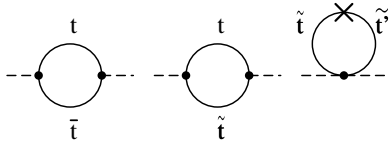,width=2in}}
\vspace*{8pt}
\caption{The cancellation of the quadratic contributions to Higgs mass square at one
loop in the top quark sector.\label{fermionloop}}
\end{figure}

In the fermion sector, to cancel the top loop contribution to the 
radiative corrections to the Higgs boson  mass, one needs to 
introduce a vector-like pair of the color triplet and 
iso-singlet heavy tops, $\tilde{t}$ and $\tilde{t}^{\prime}$. 
The field $\tilde{t}$ then form a triplet, together with 
$(b \; t)$, under the $SU(3)_{1}$ global symmetry,
$\chi^{T} = \left(b \; t \; T \right)$. 
The Yukawa interactions take the following form,
\begin{equation}
\mathcal{L}_{\mbox \tiny{Yuk}} =   
\frac{1}{2} \lambda_{1} f \epsilon_{ijk}\epsilon_{xy} \chi_{i}
 \Sigma_{jx} \Sigma_{ky} u_{3}^{\prime} 
+\lambda_{2} f \tilde{t} \tilde{t}^{\prime} + h.c. \; .
\end{equation}
The first term in this Yukawa Lagrangian preserves the $SU(3)_{1}$ 
global symmetry and breaks the $SU(3)_{2}$ global symmetry, while
 the mass term of the vector-like quarks preserves the $SU(3)_{2}$ 
and breaks $SU(3)_{1}$. 
Due to the $SU(3)_{1}$ global symmetry, the couplings of $t_{3}$ to 
$h^{0} u_{3}^{\prime}$ and $\tilde{t}u_{3}^{\prime}$, and the quartic 
coupling $h^{0} h^{0\ast} \tilde{t}u_{3}^{\prime}$ are related,
\begin{equation}
\mathcal{L}_{\mbox{\tiny Yuk}} \rightarrow -i \lambda_{1} \biggl[
\sqrt{2} h^{0} t_{3} + i f \tilde{t} - i h^{0} h^{0\ast}
 \frac{\tilde{t}}{f} \biggr] u_{3}^{\prime} + h.c. \; .
\end{equation}  
These relations lead to the cancellation of the quadratic divergences
 among diagrams shown in Fig.~\ref{fermionloop} in the fermion sector at one-loop. 

Because of the non-abelian transformation, $h \rightarrow h + \epsilon$,
 the quadratic contributions to $m_{H}$ can only arise at the two-loop 
level, due to the following scalar and fermion interactions, 
\begin{eqnarray}
\mathcal{L}_{s} & = & 
\frac{a^{2}}{2} f^{4} \bigg\{ g_{j}^{2} \sum_{a}
\mbox{Tr} \biggl[ \biggl(Q_{j}^{a}\Sigma\biggr) + \biggl(Q_{j}^{a}\Sigma\biggr)^{\ast}
\biggr] + g_{j}^{\prime 2} \mbox{Tr} 
\biggl[ \biggl(Y_{j} \Sigma\biggr) 
+ \biggl( Y_{j} \Sigma\biggr)^{\ast}\biggr]\bigg\}
\\
\mathcal{L}_{f} & = & 
-\frac{a^{\prime}}{4} \lambda_{1}^{2} f^{4} 
\epsilon^{wx} \epsilon_{yz}\epsilon^{ijk} \epsilon_{kmn}
\Sigma_{iw}\Sigma_{jx}\Sigma^{\ast my}\Sigma^{\ast nz}
\; ,
\end{eqnarray}
where the parameters $a$ and $a^{\prime}$ are the coefficients that 
parametrize the unknown UV physics.
The scalar potential arises by integrating out the heavy top,
 $T$, as well as the heavy gauge bosons,
\begin{equation}
V_{\mbox{\tiny CW}} = \lambda_{\Phi^{2}} f^{2} \mbox{Tr}
 (\Phi^{\dagger}\Phi) + \lambda_{H\Phi H} f (H\Phi^{\dagger} H)
 - \mu^{2} HH^{\dagger} + \lambda_{H^{4}} (HH^{\dagger})^{2} \; ,
\end{equation}
where 
\begin{eqnarray}
\mu^{2} & \sim & a \frac{f^{2}}{16\pi^{2}}
\\
4 \lambda_{H^{4}} & = & \lambda_{\Phi^{2}} = \frac{a}{2} \biggl[
\frac{g^{2}}{s^{2}c^{2}} + \frac{g^{\prime 2}}{s^{\prime 2} c^{\prime 2}} \biggr] 
+ 8 a^{\prime} \lambda_{1}^{2}
\\
\lambda_{H\Phi H} & = & -\frac{a}{4} 
\biggl[ \frac{g^{2}(c^{2}-s^{2}}{s^{2}c^{2}} + 
\frac{g^{\prime 2}(c^{\prime 2}-s^{\prime 2})}{s^{\prime 2} c^{\prime 2}} \biggr]
 + 4 a^{\prime} \lambda_{1}^{2} \; .
\end{eqnarray}

The complete Feynman rules of the littlest Higgs model has been presented in 
Ref.~\refcite{Han:2003wu}. 
Various collider phenomenology of the littlest Higgs model has been 
discussed extensively in Ref.~\refcite{Han:2003wu,Burdman:2002ns}, and the
 flavor sector in this model has been studied in 
Ref.~\refcite{Huo:2003vd}, including the generation of fermion
 masses\cite{Bazzocchi:2003ug}.

\subsubsection{Littlest Higgs with T-parity}

As we will see in the next section, the littlest Higgs model
 is severely constrained by the precision electroweak data.
 The most stringent one comes from the tree level contributions 
to the $\rho$ parameter, due to the presence of the $W_{L}W_{H} HH$ 
coupling and the tri-linear $H^{T}\Phi^{\dagger} H$ coupling, which breaks the tree
 level custodial symmetry explicitly. In Ref.~\refcite{Cheng:2003ju}, 
Cheng and Low found that these operators can be forbidden by 
imposing a discrete $Z_{2}$ symmetry, called the T-parity.
 Under the T-parity, all new particles, except the heavy top partner,
 $\tilde{t}$,  that are responsible for canceling the SM contributions
 to the one-loop quadratic divergences in the Higgs potential are odd, 
while all the other particles are
 T-even\footnote{See also Ref.~\refcite{Cheng:2005as}.}. 
As a result, the heavy particle contributions to the observables 
involving only the SM particles in the external states are forbidden at 
the tree level. These contributions of the new particles to precision 
EW observables can arise only at the loop levels in this model, and 
thus the constraints in this model is not as stringent as in the littlest Higgs 
model. 
Due to T-parity, all T-odd particles have to be pair-produced, 
and the lightest stable particle can be the candidate of the
weak scale dark matter\cite{Hubisz:2004ft}.  Various collider phenomenology and 
flavor constraints in this model has been discussed 
in Ref.~\refcite{Hubisz:2005tx} and \refcite{Hubisz:2005wy}, respectively. 
It was found that the collider signatures in this model mimic that of the 
MSSM.


\subsubsection{$SU(6)/Sp(6)$ Little Higgs Model}

In Ref.~\refcite{Low:2002ws}, the global symmetry of the 
model is chosen to be $SU(6)$ which is broken down to $Sp(6)$. 
This has the advantage that the pseudo-Goldstone multiplet does not 
contain a SU(2) triplet component. This thus weakens the constraints from 
precision data.  
The gauged subgroup of $SU(6)$ is $[SU(2)\times U(1)]^{2}$, 
which is broken down to its diagonal subgroup, 
$[SU(2)\times U(1)]_{\mbox{\tiny SM}}$.   
There are fourteen Goldstone bosons $(35-21=14)$ resulting 
from the global symmetry breaking. Four of them are eaten due
 to gauge symmetry breaking. Thus at low energy spectrum, there
 are two complex doublet scalar fields. At the TeV scale, there are 
one complex singlet scalar, a few pairs of vector-like colored fermions 
and an extra copy of the $SU(2)\times U(1)$ gauge bosons. 
By turning off either one of the SU(2) gauge coupling constants, there 
is an enhanced SU(4) global symmetry which forbids a square mass term for the Higgs.  
Thus the quadratic divergences to the Higgs mass can only arise at two loop. 
The issue of vacuum stability 
in the presence of the anti-symmetric condensate of 
the $\Sigma$ field in this model has also been 
investigated\cite{Low:2002ws}.

\subsection{Models with Simple Group} 

\subsubsection{Little Higgs Model from A Simple Group}

In the model proposed in Ref.~\refcite{Kaplan:2003uc}, 
the global symmetry is $[SU(4)]^{4}$ which breaks down to 
$[SU(3)]^{4}$. And the gauge subgroup is $SU(4)\times U(1)$ 
which breaks down to $SU(2)\times U(1)$. There are $(15-8)\times 4 = 28$ 
NBG's: 12 of them are eaten due to gauge symmetry breaking and the 
remaining 16 real components are decomposed into two complex doublets, 
three complex $SU(2)$ singlets and two real singlet scalars. By embedding 
the electroweak $SU(2)_{L}$ gauge group into a simple group such as $SU(3)$ 
or SU(4), this model has a nice feature that 
the cancellation of one-loop quadratic divergences from the gauge 
and perturbatively coupled fermion loops is automatic. Generation of neutrino masses in this 
model has been discussed in Ref.~\refcite{Lee:2005mb}.

\subsubsection{A Simple Model of Two Little Higgses}

The coset of the model proposed in Ref.~\refcite{Skiba:2003yf} is 
$SU(9)/SU(8)$ while the electroweak gauge symmetry of this model 
is expanded to $SU(3)\times U(1)$ which is embedded into $SU(9)$. 
Due to the enlarged gauge group, $SU(3)\times U(1)$, there is no 
mixing induced by the VEV of the Higgs boson between the light 
and the heavy gauge bosons in this model, and it has only one 
additional gauge boson, the $Z^{\prime}$.

\subsection{Moose Models}

\subsubsection{Minimal Moose Model}

In moose models, the electroweak sector of the SM is embedded into a theory  
with a product global symmetry, $G^{N}$, which is broken by a set of condensates 
transforming as bi-fundamental representations under $G_{i} \times G_{j}$ for 
pair (i,j). Some subgroup of $G^{N}$ which contains the SM is being gauged for each site; 
it is then broken down to the SM gauge group $SU(2)\times U(1)$ by the bi-fundamental condensate 
at the TeV scale. The minimal moose model proposed in Ref.~\refcite{Arkani-Hamed:2002qx} 
has $[SU(3)]^{8}$ as global symmetry. 
The gauge subgroup of the model is $[SU(3)\times SU(2)\times U(1)]$. 
The spectrum at low energy contains two complex Higgs doublets, a complex triplet and a singlet.

\subsubsection{$SO(5)$ Moose Model}

By enlarging the global group in the minimal moose model to 
$[SO(5)]^{8}=[SO(5)_{L}]^{4}\times[SO(5)_{R}]^{4}$ and the
 gauge group to $SO(5)\times SU(2)\times U(1)$, it  has 
been shown in Ref. \refcite{Chang:2003un} that the model 
preserves an approximate custodial $SU(2)$ symmetry. As a generic feature 
of the moose models, it is a two-Higgs doublet model at low energy. At the TeV scale, 
there is a colored Dirac fermion, a triplet Higgs and extra gauge bosons in the spectrum.

\subsubsection{$SO(9)$ Moose Model}

The mass splitting among the triplet components give 
large contributions to the $\rho$ parameter, which 
occurs in the $SO(5)$ moose model. This problem can 
be alleviated by expanding the global symmetry to 
$SO(9)$ as shown in Ref.~\refcite{Chang:2003zn}. In the $SO(9)$ 
moose model, the coset is $SO(9)/[SO(5)\times SO(4)]$, while the 
embedded gauge symmetry is $SU(2)_{L}\times SU(2)_{R}\times SU(2)\times U(1)$.
At the TeV scale, there are three triplet Higgses present whose VEVs preserve 
an approximate custodial symmetry. This is an essential feature that reduces the contributions 
to the $\rho$ parameter from the scalar sector in this model.

\section{Precision Electroweak Constraints}\label{preew}

As the electroweak sector of the SM has been tested to a 
very high accuracy, an important test of the validity of any 
new models is through the agreement between the predictions 
of these models with the precision data. A SM-like renormalization 
procedure with three input parameters in the gauge sector is valid 
as long as the models has tree level custodial symmetry and thus $\rho =1$ 
at three level. Examples of models with $\rho =1 $ at tree level include 
the SM augmented by additional Higgs doublets or singlets, additional 
fermion families and MSSM. On the other hands, in many extension of the SM, 
the tree level custodial symmetry is no longer a good symmetry of 
the model, {\it i.e.} $\rho \ne 1$ already at the tree level.
Models of this type include the left-right symmetric model based on 
$SU(2)_{L}\times SU(2)_{R}\times U(1)_{B-L}$, models with additional 
$SU(2)_{L}$ triplet Higgses (which might be relevant to generation 
of neutrino masses\cite{Chen:2000fp,Chen:2003zv}), and various little 
Higgs models, to name a few.

In the gauge sector of the SM, there are only three {\it independent} 
parameters, the $SU(2)_{L}$ and $U(1)_{Y}$ gauge coupling constants, 
$g$ and $g^{\prime}$, as well as the VEV of the Higgs doublet, $v$. 
Once these three parameters and their counter terms are fixed by 
the experimental data, all other physical observables in the gauge 
sector can then be predicted in terms of these three input 
parameters\footnote{In addition to these three input parameters in the 
gauge sector, there are additional input parameters in the fermion and 
scalar sectors. These can be chosen to be the fermion and scalar masses.}. 
A special feature of the SM with the assumption of one Higgs doublet 
is the validity of the tree level relation, $\rho = 1 = 
\frac{M_{W}^{2}}{M_{Z}^{2} c_{\theta}^{2}}$ due to the tree level 
custodial symmetry. There is thus a definite relation between the 
W-boson mass and the Z-boson mass. 
Of course, one can equivalently choose any three physical observables 
as the input parameters in the gauge sector. If we choose $G_{\mu}, \, 
M_{Z}$ and $\alpha$ as the three input parameters in the gauge sector,
 the W-boson mass, $M_{W}$, then is predicted in the usual way via muon-decay,
\begin{equation}
M_{W}^{2} = \frac{\pi \alpha}{\sqrt{2} G_{\mu} s_{\theta}^{2}} \biggl[ 1 
+ \Delta r \biggr] \; ,
\end{equation}
where $\Delta r$ summarizes the one-loop radiative corrections, 
and it is given in terms of the gauge boson self-energy two point functions as,
\begin{eqnarray}
\Delta r & = & -\frac{\delta G_{\mu}}{G_{\mu}} - 
\frac{\delta M_{W}^{2}}{M_{W}^{2}} + \frac{\delta \alpha}{\alpha}
 - \frac{\delta s_{\theta}^{2}}{s_{\theta}^{2}}
\\
& = & \frac{\Pi^{WW}(0)-\Pi^{WW}(M_{W})}{M_{W}^{2}} + \Pi^{\gamma\gamma
 \, \prime}(0)  + 2 \frac{s_{\theta}}{c_{\theta}} 
\frac{\Pi^{\gamma Z}(0)}{M_{Z}^{2}} -  \frac{\delta s_{\theta}^{2}}{s_{\theta}^{2}}
\nonumber \; .
\end{eqnarray}
The counter term for the weak mixing angle $s_{\theta}$ which is defined 
through the W- and Z-boson mass ratio, $s_{\theta}^{2} = 1 
- \frac{M_{W}^{2}}{M_{Z}^{2}}$, is given by,
\begin{equation}
\frac{\delta s_{\theta}^{2}}{s_{\theta}^{2}} = \frac{c_{\theta}^{2}}{s_{\theta}^{2}} 
\biggl[ 
\frac{\Pi^{ZZ}(M_{Z})}{M_{Z}^{2}}-\frac{\Pi^{WW}(M_{W})}{M_{W}^{2}}\biggr]
\; .
\end{equation}
Both of the two point functions, $\Pi^{WW}(M_W)$ and $\Pi^{WW}(0)$,
 have identical leading quadratic $m_{t}$ dependence, 
$\frac{\sqrt{2}G_{\mu}}{16\pi^{2}}3m_{t}^{2}\bigl( 1 
+ 2 \ln\frac{Q^{2}}{m_{t}^{2}}\bigr)$, and thus their difference
 is only logarithmic. The two-point function, $\Pi^{\gamma\gamma\prime}(0)$, 
is also logarithmic in $m_{t}$. However, the difference between 
$\Pi^{WW}(M_{W})$ and $\Pi^{ZZ}(M_Z)$ has quadratic dependence in $m_{t}$. 
Thus the prediction for $M_{W}$ is quadratic in $m_{t}$. 
 
In the presence of a $SU(2)_{L}$ triplet Higgs, on the other hand, 
a tri-linear coupling between the doublet and the triplet Higgs, 
$H^{T}\Phi^{\dagger}H$, is allowed by the gauge symmetry $SU(2)_{L} \times U(1)_{Y}$. 
So unless one imposes a discrete symmetry to forbid such a tri-linear 
interaction, the VEV of the triplet is non-zero, 
$\left< v^{\prime} \right> \ne 0$. This thus leads to the need for
 a fourth input parameter in the gauge sector, with the fourth 
parameter being the VEV of the triplet Higgs, $v^{\prime}$. 
Many of the familiar predictions 
of the Standard Model are drastically changed by the need for this extra
input parameter\cite{Passarino:1990xx}.
One can equivalently choose the effective leptonic mixing angle, 
$s_{\theta}$, as the fourth input parameter, where $s_{\theta}$ 
is defined through the ratio of the vector to axial vector parts 
of the $Ze\overline{e}$ coupling, $4s_{\theta}^{2}-1=
\frac{\mbox{Re}(g_{V}^{e})}{\mbox{Re}(g_{A}^{e})}$. 
The counter term for $s_{\theta}^{2}$ is formally 
related to the wave function renormalizations for $\gamma$ 
and $Z$ and it is given by,
\begin{equation}
\frac{\delta s_{\theta}^{2}}{s_{\theta}^{2}}  = \mbox{Re}
\bigg\{ \frac{c_{\theta}}{s_{\theta}} \biggl[
\frac{\Pi^{\gamma Z}(M_{Z})}{M_{Z}^{2}} - 
\frac{v_{e}}{2s_{\theta}c_{\theta}}
\biggl(\frac{a_{e}^{2}-v_{e}^{2}}{a_{e}v_{e}} \Sigma_{A}^{e}(m_{e}^{2}) +
\frac{\Lambda_{V}^{Zee}(M_{Z})}{v_{e}}
-\frac{\Lambda_{A}^{Zee}(M_{Z})}{a_{e}}\biggr)\biggr]\bigg\}
\; ,\label{cs}
\end{equation}
where $\Sigma_{A}^{e}$ is the axial part of the electron 
self-energy and $\Lambda_{V,A}^{Zee}$ are the vector and 
axial vector parts of the $Ze\overline{e}$ vertex corrections. 
Contrary to the SM case in which the $m_{t}$ dependence is quadratic,   
in models with a triplet Higgs the dominant contribution in Eq.~\ref{cs}, 
$\Pi^{\gamma Z}(M_Z)$, depends on $m_{t}$ only logarithmically. 
Due to this logarithmic dependence,
 the constraint on the model is weakened. On the other hand,
 the scalar contributions become important as they are quadratic 
due to the lack of the tree level custodial symmetry, 
as pointed out in Ref.~\refcite{Passarino:1990xx,Toussaint:1978zm,Einhorn:1988tc,Chen:2003fm,Chen:2005jx}.

\subsection{Littlest Higgs Model}

In many little Higgs models, the $\rho$ parameter also differs from one 
already at the tree level, 
due to the presence of the triplet Higgs which acquires a non-vanishing 
VEV. (For analyses based on tree level constraints can be found in Ref.~\refcite{Csaki:2002qg} 
and those including the heavy top effects at one loop can be found in Ref.~\refcite{Hewett:2002px}.)  
 A consistent renormalization scheme  thus requires an additional
 input parameter in the gauge sector.
  In Ref.~\refcite{Chen:2003fm}, 
the authors choose 
the muon decay constant $G_{\mu}$, the physical Z-boson mass $M_{Z}^{2}$, 
the effective lepton mixing angle $s_{\theta}^{2}$ and the fine-structure 
constant $\alpha(M_{Z}^{2})$ as the four independent input parameters in 
the renormalization procedure. The $\rho$ parameter, defined as,
$\rho \equiv M_{W_{L}}^{2}/(M_{Z}^{2}c_{\theta}^{2})$, 
and the $W$-boson mass, which is defined through muon decay, 
are then derived quantities.
Since the loop factor occurring in radiative corrections,  
$\frac{1}{16\pi^2}$, is similar in magnitude to the expansion parameter, 
$\frac{v^{2}}{f^{2}}$, of the chiral perturbation 
theory for $f$ of a few TeV, the one-loop radiative corrections can
be comparable in size to  the next-to-leading order contributions at tree 
level. Both types of corrections are of the order of a few percent.

\begin{figure}[htb]
\psfrag{M(theory)-M(exp) (GeV)}[][]{$M_{\mbox{theory}}
-M_{\mbox{exp}}$ (GeV)}
\psfrag{x_L}[][]{$x_{L}$}
\psfrag{Mw  }[][]{$\delta M_{W_{L}}$(total) $\qquad $}
\psfrag{Mz  }[][]{$\delta M_{Z}$(total) $\qquad $}
\psfrag{Mwtree  }[][]{$\delta M_{W_{L}}$(tree) $\quad $}
\psfrag{deltaMw}[][]{1 $\sigma$ limit on 
$\delta M_{W_{L}}$ (exp) $\qquad \qquad \qquad$ }
\psfrag{deltaMz}[][]{1 $\sigma$ limit on 
$\delta M_{Z}$ (exp) $\quad \qquad \qquad \qquad $}
\centerline{\psfig{file=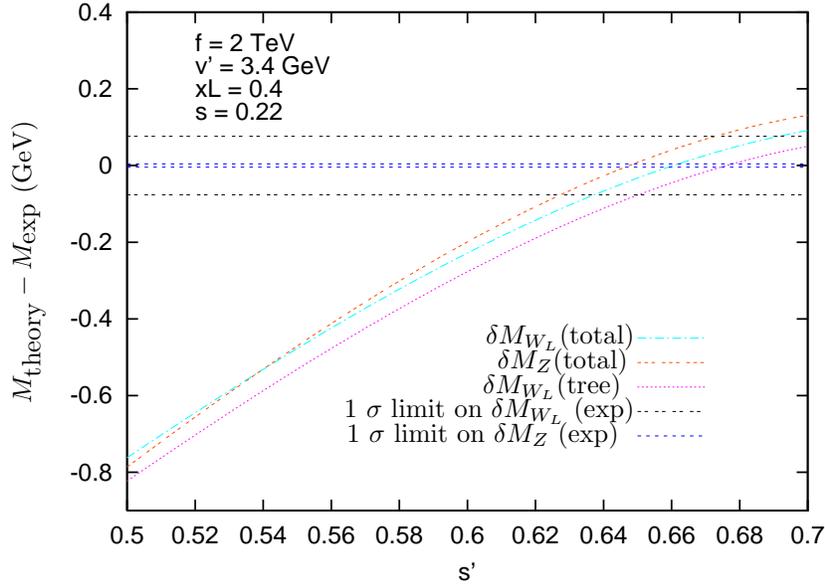,width=3in,angle=270}}
\vspace*{8pt}
\caption{%
Prediction for $M_{W_{L}}$ as a function of the mixing angle 
$s^\prime$ at the tree level and the one-loop level. 
Also plotted is the correlation between $M_{Z}$ and 
$s^\prime$ for fixed $s$, $v^{\prime}$ and $f$. 
The cutoff scale $f$ in this plot is $2$ $TeV$, the $SU(2)$ triplet VEV 
$v^\prime = 3.4 \; GeV$, the mixing angle $s=0.22$, and $x_{L}=0.4$. (Figure 
taken from Ref.~31)
}
\label{fig0}
\end{figure}

The effective leptonic mixing angle is defined through the ratio of the 
vector to axial vector parts 
of the $Zee$ coupling, 
\begin{equation}
4s_{\theta}^{2} -1 = \frac{\mbox{Re}(g_{V}^{e})}{\mbox{Re}(g_{A}^{e})},
\end{equation}
which differs from the naive definition of the Weinberg angle in the
 littlest Higgs model, 
$s_{W}^{2} = g^{\prime 2}/(g^{\prime 2} + g^{2})$, by,
\begin{equation}
\Delta s_{\theta}^{2} \equiv s_{W}^{2} - s_{\theta}^{2} 
 = -\frac{1}{2\sqrt{2} G_{\mu} f^{2}}
\left[ s_{\theta}^{2} c^{2} (c^{2} - s^{2}) 
- c_{\theta}^{2} (c^{\prime 2} - s^{\prime 2}) (-2 + 5 c^{\prime 2})
\right] \; .
\end{equation}
The W-boson mass is defined through muon decay,
\begin{equation}
M_{W}^{2} = \frac{\pi \alpha}{\sqrt{2} G_{\mu} s_{\theta}^{2}}
\left[ 1 + \Delta r_{\mbox{tree}} + \Delta r^{\prime} \right] \; ,
\end{equation}
where $\Delta r_{\mbox{tree}}$ summarize the tree level corrections due to the 
change in definition in the weak mixing angle as well as the 
contributions from exchange 
of the heavy gauge bosons,
\begin{equation}
\Delta r_{\mbox{tree}} = -\frac{\Delta s_{\theta}^{2}}{s_{\theta}^{2}} + 
\frac{c^{2} s^{2}}{\sqrt{2} G_{\mu} f^{2}} \; .
\end{equation}  
The one-loop radiative corrections are collected in $\Delta r^{\prime}$,
\begin{eqnarray}
\Delta r^{\prime} & = & -\frac{\delta G_{\mu}}{G_{\mu}} - 
\frac{\delta M_{W}^{2}}{M_{W}^{2}}
+ \frac{\delta \alpha}{\alpha} - \frac{\delta s_{\theta}^{2}}{s_{\theta}^{2}}
\\
& = & \frac{1}{M_{W}^{2}} \left[ \Pi^{WW}(M_{W}) - \Pi^{WW}(0) \right]
+ \Pi^{\gamma\gamma}(0)^{\prime} - 
\frac{c_{\theta}}{s_{\theta}} \frac{\Pi^{\gamma Z}(M_{Z})}{M_{Z}^{2}}
\nonumber\; .
\end{eqnarray} 
The predictions for $M_{W_{L}}$ with and without the 
one-loop contributions for $f=2$ TeV is given in Fig.~\ref{fig0}, 
which demonstrates that a low value of $f$ ($f\sim 2$ TeV) is allowed by 
the experimental restrictions from the $W$ and $Z$ boson masses,\cite{Chen:2003fm} due to 
cancellations among the tree-level and one-loop corrections. 
This shows the importance of a full one-loop calculation in placing the electroweak 
precision constraints.

\subsection{Constraints in Other Little Higgs Models}

Except for the littlest Higgs model with T-parity, where $f$ can be as low as 500 GeV\cite{Hubisz:2005tx},  
all other models discussed in the previous section receive tree level contributions to the EW observables. 
The precision EW constraint in the $SU(6)/Sp(6)$ model has been analyzed in 
Ref. \refcite{Gregoire:2003kr}. Due to the absence of the triplet Higgs, the contribution 
to the $T$ parameter is minimal, and the scale for $f$ in this model can be 1 TeV\cite{Gregoire:2003kr}.
In the simple group model based on $[SU(4)]^{4}/[SU(3)]^{4}$, the precision EW fit has been performed in 
Ref.~\refcite{Csaki:2003si}, and the bound on $f$ is found to be 4.2 TeV. In the $SU(9)/SU(8)$ 
simple group model, the bound on $f$ is slightly improved to be 3.3 TeV~\cite{Skiba:2003yf,Marandella:2005wd}.
The precision EW constrains in the minimal moose model have been investigated 
in Ref.~\refcite{Kilic:2003mq}, in which $f$ is found to be very severely constrained. 
Due to the approximate custodial symmetry in the SO(5) and SO(9) moose models, this bound can be  
relaxed\cite{Casalbuoni:2003ft}.

\section{Other Issues}\label{other}


It has been pointed out that in many new models with Physics beyond 
 the Standard Model, there is generally an implicit fine-tuning 
among the model parameters which may be over-looked at first
 glance but show up in a systematic analysis\cite{Barbieri:1987fn}. 
Such implicit fine-tuning, which is needed to 
render the models phenomenologically viable, 
has been quantified by Barbieri and Giudice in Ref.~\refcite{Barbieri:1987fn}. 

\begin{figure}[b!]
\centerline{\psfig{file=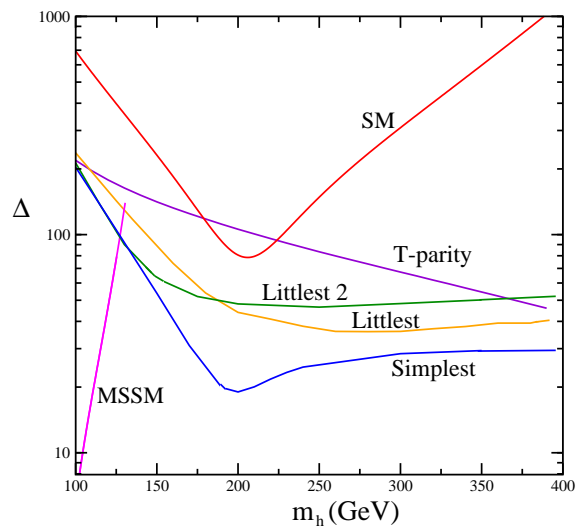,width=3in,angle=270}}
\vspace*{8pt}
\caption{\label{dvsmh}Comparison of fine-tuning in SM (with a cutoff of 10 TeV), MSSM and 
various Little Higgs models. (Figure taken from Ref. 39)}
\end{figure}

In Ref.~\refcite{Casas:2005ev}, Casas, Espinosa and Hidalgo
 examine this issue in the little Higgs models and find that
 the degree of such implicit fine-tuning is usually much more
 substantial than the rough estimate.      
If we define the amount of fine-tuning in the Higgs VEV  
in the model {\it a la}
Barbieri and Giudice\cite{Barbieri:1987fn} as, $v^{2} =
 v^{2}(p_{1},p_{2},...)$, where $p_{i}$ for $(i=1,....)$ are input parameters
 of the model, then the amount of fine-tuning associated with 
parameter $p_{i}$ is given by,
\begin{equation}
\frac{\delta M_{Z}^{2}}{M_{Z}^{2}} = 
\frac{\delta v^{2}}{v^{2}} = \Delta_{p_{i}}\frac{\delta p_{i}}{p_{i}}
\; ,
\end{equation} 
and the total amount of fine-tuning in the model is defined as,
\begin{equation}
\Delta \equiv \sqrt{\sum_{i} \Delta_{p_{i}}^{2}} \; .
\end{equation}
Fig.~\ref{dvsmh} shows the fine-tuning that exists 
in various versions of the little Higgs models as well as in the SM and MSSM.
 Such implicit fine-tuning in the little Higgs models has been found
 to be more severe compared to the fine-tuning required in MSSM\cite{Casas:2005ev}. 
(See also Ref.~\refcite{Bazzocchi:2005gs}.)


Since the little Higgs models are effective theories valid only up 
to the cut-off scale, $\Lambda \sim 
4\pi f \sim 10$ TeV, an important question one has to address is what
 lies beyond this cut-off scale\cite{Chang:2003vs}.  
A few possible UV completions have been speculated.
 In Ref.~\refcite{Katz:2003sn}, the little Higgs model
 is UV completed into a model in which the little Higgs
 is composite at 10 TeV and the model has a matter parity,
 $(-1)^{(2S+3B+L)}$, similar to the R-parity in SUSY. 
In Ref.~\refcite{Kaplan:2004cr} and \refcite{Batra:2004ah},
 the little Higgs model is completed by another little 
Higgs model. This thus postpones the onset of the strong 
coupling regime to $\sim 100$ TeV.  In Ref.~\refcite{Thaler:2005en}
 an alternative was proposed in which the littlest Higgs model is
 UV completed by a five-dimensional Anti-de Sitter space where
 the global SU(5) symmetry in 4D littlest Higgs model
 corresponds to an SU(5) gauge symmetry in the 5D bulk.  

The issue of vacuum stability in  various little Higgs models 
has been investigated in 
Ref.~\refcite{Datta:2004td}, while their finite temperature 
effects have been studied in 
Ref.~\refcite{Trodden:2004ea}. An attempt to supersymmetrize
 the little Higgs 
model has been made in Ref.~\refcite{Csaki:2005fc}, and it was
 found that the model can be embedded into an $SU(6)$ GUT.

\section{Conclusions}\label{concl}

Little Higgs models are a new approach to stabilizing the Higgs
 mass. The Higgs boson in these models arises as a pseudo-Nambu-Goldstone 
boson resulting from the spontaneous breaking of some global symmetry. 
No interaction in these models alone can break the 
complete global symmetries that prohibit the quadratic contribution 
to the Higgs potential. Thus only when these interactions act 
collectively can the complete global symmetries be broken to give  
a quadratic mass term to the Higgs boson. The quadratically divergent   
contributions to the Higgs potential can therefore arise only at the 
two-loop level. In this note, I review  various little Higgs models and 
the precision EW constraints placed in them. In most of these models, the parameter 
space is severely constrained by the precise electroweak data, 
with the exception of the littlest Higgs model with T-parity, in which a low cutoff 
scale is allowed. The importance of a full one-loop calculation is also emphasized 
and demonstrated explicitly in the case of the littlest Higgs model. 
There are other important questions such as the UV completion, 
vacuum stability and 
the contributions to the electroweak precision observables from 
higher dimensional operators at the cutoff scale, which are either out of the scope of 
this review or have only been briefly mentioned. 
More careful studies on these issues are clearly needed.

\section*{Acknowledgments}
I thank my collaborators, S. Dawson, T. Krupovnickas and K.T. Mahanthappa  
for numerous discussions.  The hospitality 
of the Aspen Center for Physics, where part of this work was 
done, is also acknowledged. Fermilab is operated by Universities 
Research Association Inc. under Contract No. DE-AC02-76CH03000 
with the U.S. Department of Energy.

\end{document}